\documentclass[aps,twocolumn]{revtex4}
\usepackage{graphicx}
\usepackage{amsmath}
\usepackage{amssymb}
\newcommand{\be}{\begin{equation}}
\newcommand{\ee}{\end{equation}}
\newcommand{\bea}{\begin{eqnarray}}
\newcommand{\eea}{\end{eqnarray}}

\begin{document}
\title{\Large \bf Cosmological constant, brane tension and
large hierarchy in a generalized Randall-Sundrum braneworld scenario}
\author{Saurya Das$^1$ \footnote{E-mail: saurya.das@uleth.ca },
Debaprasad Maity$^2$ \footnote{E-mail: debu@imsc.res.in} and
Soumitra SenGupta$^3$ \footnote{E-mail: tpssg@iacs.res.in}}
\affiliation{$^1$Department of Physics,  University of Lethbridge\\
4401 University Drive, Lethbridge, Alberta - T1K 3M4, Canada\\
$^2$The Institute of Mathematical Sciences,\\
CIT Campus, Chennai - 600 113, India\\
$^3$Department of Theoretical Physics and Centre for
Theoretical Sciences,\\
Indian Association for the Cultivation of Science,\\
Kolkata - 700 032, India}

\begin{abstract} We consider a generalized Randall Sundrum (RS)
brane world scenario with a cosmological constant $\Omega$ induced
on the visible brane. We show that for $\Omega<0$, resolution of
the hierarchy problem requires an upper bound on the magnitude of
$\Omega$. The corresponding tension on the visible brane can be
positive or negative. On the other hand, there is no such bound
for $\Omega>0$. However, in this case, the resolution of the
hierarchy problem along with the tuning of the value of the cosmological constant to its observed
value closed to $+10^{-124}$ (in Planck units) naturally lead to the tuning of the modulus to
a small value of inverse Planck length as estimated in the original RS scenario.

%
\end{abstract}

\maketitle

Some of the intriguing questions about our physical universe,
which remain unanswered, are: \\
(i) Why does it appear to have $(3+1)$ space-time
dimensions? Are there additional unobserved dimensions? \\
(ii) Why is the ratio of the electroweak scale/Higgs mass ($m$) to
the Planck mass ($m_0$) so tiny ($\simeq 10^{-16}$)? This gives
rise to the {\it gauge hierarchy problem}.\\
(iii) Why is the observed value of the cosmological constant
$\Omega$ extremely small ($\Omega \simeq 10^{-124}$) (in Planck
units)? This gives rise to the {\it cosmological fine tuning
problem}.

In the so-called brane world models proposed during the last
decade \cite{RS,ADD,cohen}, it was shown that questions (i) and
(ii) may be related, in the sense that if one assumes that the
space-time dimension exceeds four, the hierarchy problem can be
solved. A lot of work has also been done recently in an attempt to
relate the questions (i) and (iii) by adopting various approaches
such as the domain wall scenarios \cite{rubakov,verlinde} and self
tuning mechanism in large extra dimensions
\cite{silverstein,carroll,csaki}. While the RS two-brane model is
particularly successful in resolving the fine tuning problem
without bringing in any arbitrary intermediate scale between the
Planck and the Tev scale, it has  a somewhat unsatisfactory but
inevitable feature of having a negative tension visible brane to
describe our Universe. It has been shown that such negative
tension branes are intrinsically unstable. Furthermore the
effective visible 3-brane being flat has zero cosmological
constant which is not consistent with its presently observed small
value. In this article we extend such warped geometric model to
include a non-zero cosmological constant and look for a possible
positive tension Tev brane when a large hierarchy exists between
the two branes. A motivation to look for positive tension branes
lies in string inspired brane world scenarios, in which the
relevant D-branes have positive tension.

Here by generalizing the RS model to include a non-vanishing
cosmological constant on the visible brane, we show that questions
(i), (ii) and (iii), as well as the issue of brane tensions are
intimately related.
We demonstrate that while the regime of positive cosmological
constant on the visible 3-brane (de-Sitter) strictly implies a
negative brane tension, that with negative cosmological constant
(anti de-Sitter) admits of both positive and negative tensions of
the visible brane. For both the regions, the desired warping from
Planck to Tev scale can be achieved as a proper resolution of the
gauge hierarchy problem. However larger is the magnitude  of the 4d cosmological
constant ( + ve or - ve) further away is the value of the modulus from the Planck length leading
to a new hierarchy of scales. This obviously brings back the fine tuning problem in a new guise and
is undesirable. 

In the RS scenario, it was proposed
that our universe is five dimensional, described by the metric
\cite{RS}:
\be ds^2 = e^{-2 k r y} \eta_{\mu\nu} dx^{\mu}dx^{\nu} + r^2 dy^2
\ee
where Greek indices $\mu,\nu,\dots$ run over $0,1,2,3$ and refer
to the $4$ observed dimensions, $y$ signifies the coordinate on
the additional spacelike dimension of length $r$, $\Lambda$ is the
bulk cosmological constant, $ k \equiv \sqrt{- \Lambda/{12
M^3}}\approx $ Planck mass. The factor $e^{-2 k r y}$ is known as
the warp factor. The geometry of the extra dimension is orbifolded
by $S^1/Z_2$. The constant $y$ slices at $y=0$ and $y=r \pi$ are
known as the {\it hidden} and {\it visible} branes, the observable
universe being identified with the latter which has a negative
brane tension as opposed to the {\it hidden} brane which has a
positive brane tension.
It can be shown that even if the Higgs (or any other) mass
parameter in the five-dimensional Lagrangian is of the order of
Planck scale $m_0$ ($\approx 10^{16}~TeV$), on the visible
four dimensional brane, it gets `warped' by a factor of the form:
\be m = m_0  e^{-2 k r \pi}~. \ee
For, $ k r\simeq 11.84$, one gets $m \approx 1~TeV$. Thus, in this
picture, the origin of a small Higgs mass lies in the warped
geometry of five dimensional spacetime.

In \cite{RS}, it was also shown that the cosmological constant
induced on the visible brane is zero.
In this paper we demonstrate that the last condition can be
relaxed for a more general warp factor, such that the metric is
given by:
\be ds^2 = e^{-2A(y)} g_{\mu\nu} dx^{\mu}dx^{\nu} + r^2 dy^2~.
\label{ourmetric} \ee
For the above metric  the visible brane can have a negative or a positive
cosmological constant. Defining $e^{-A(kr\pi)}=m/m_0 = 10^{-n}$ and the
magnitude of the induced cosmological
constant $= 10^{-N}$ (in Planck units), we show that for negative cosmological constant
$N$ cannot be less
than a minimum value given by $N_{min} = 2n$. This implies a very small upper bound of the magnitude of the
cosmological constant. Although for positive value of
$\Omega$ no such bound exists in general,
the need for resolution of the hierarchy problem without
introducing a new scale in the theory ( i.e keeping the value of the modulus close to Planck length ) , restricts the
cosmological constant to be very very small. The
corresponding brane tension for both Tev and the Planck branes
are determined
for these two different scenarios.

We start with the metric (\ref{ourmetric}) and evaluate the
function $A(y)$ which extremises the action:
\begin{equation}
S = \int d^5x \sqrt{-G} ( M^3 {\cal R} - \Lambda) + \int d^4x
\sqrt{-g_i} {\cal V}_i
\end{equation}
where $\Lambda$ is the bulk cosmological constant, ${\cal R}$ is
the bulk ($5$-dimensional) Ricci scalar and ${\cal V}_i$ is the
tension of the $i^{th}$ brane ($i=hid (vis)$ for the hidden
(visible) brane). Note that $g_{\mu\nu}$ is the four dimensional
metric.


The resulting Einstein equations are:
\bea
\label{equation} {^4G_{\mu\nu}} - g_{\mu\nu} e^{-2 A}[-6 A'^2
+ 3 A'']
&=& -
\frac{\Lambda}{2M^3}g_{\mu\nu} e^{-2 A} \\
-\frac 1 2 e^{2 A}~{^4R} + 6 A'^2 &=& -\frac {\Lambda}{2M^3}
\label{equation2} \eea
with the boundary conditions \be \label{boundary} [A'(y)]_i =
\frac{\epsilon_i}{12 M^3} {\cal V}_i ~, \ee
where $\epsilon_{hid}=-\epsilon_{vis}=1$. In the above,
${^4G_{\mu\nu}}$ and $^4 R$ are the four dimensional Einstein
tensor and Ricci scalar respectively, defined with respect to
$g_{\mu\nu}$. Dividing both sides of Eq.(\ref{equation}) by
$g_{\mu\nu}$, for any $\mu,\nu$, and rearranging terms, it is seen
that one side contains $A(y)$ and its derivatives, depending on
the extra coordinate $y$ alone, while the other side depends on the
brane coordinates $x^\mu$ alone \cite{verlinde,gherghetta}. Thus
each side is equal to an arbitrary constant, $\Omega$ say. Thus,
we get from Eq.(\ref{equation2})
\footnote{Our Eq.(\ref{eom1}) is equivalent to Eq.(17) of T.
Shiromizu, K. Maeda, M. Sasaki, Phys. Rev. {\bf D62} (2000)
024012, in the absence of matter on the brane. }
 :
\bea  {^4 G_{\mu\nu}} = -\Omega g_{\mu\nu}~\label{eom1}~~&&, \\
e^{-2A}~\left[  -6 A'^2 + 3 A'' - \frac{\Lambda}{2M^3} \right] &=&
-\Omega~. \label{eom2} \eea
Computing $^4 R$ from Eq.(\ref{eom1}), and substituting in
Eq.(\ref{equation2}), $A'$ can be evaluated, which on further
substitution in Eq.(\ref{eom2}), yields a simplified expression
for $A''$:
\bea 6 A'^2 &=& - \frac{\Lambda}{2M^3} + 2 \Omega e^{2A}
~\label{eom3} \\
3 A'' &=& \Omega e^{2 A}~.\label{eom4} \eea
The above corresponds to a constant curvature brane spacetime, as
opposed to a Ricci flat spacetime, which is normally assumed. For
example, for $\Omega > 0$ and $\Omega < 0$, $g_{\mu\nu}$ may
correspond to dS-Schwarzschild and AdS-Schwarzschild spacetimes
respectively.

For Ads bulk i.e.  $\Lambda<0$,
we first consider the regime for which the induced cosmological constant $\Omega$ on the visible brane is
negative. Defining the parameter $\omega^2 \equiv -\Omega/3k^2 \geq 0$, we
get the following solution for the warp factor, satisfying
Eqs.(\ref{eom3}-\ref{eom4}) :
\be e^{-A} = \omega \cosh\left(\ln \frac {\omega} c_1 + ky \right)
\ee
Note that the RS solution $A=ky$ is recovered in the limit $\omega
\rightarrow 0$.
>From this and Eq.(\ref{boundary}), the brane tensions follow:
%
%
\begin{equation}
{\cal V}_{vis} =  12 M^3 k \left[\frac {\frac { \omega^2} {c_1^2}
e^{ 2 k r \pi} - 1} { \frac { \omega^2} {c_1^2} e^{ 2 k r \pi} +
1}\right]~ ;~ {\cal V}_{hid} =  12 M^3 k \left[\frac {1 - \frac {
\omega^2} {c_1^2}} { 1 + \frac { \omega^2} {c_1^2}}\right]
\label{boundary1}
\end{equation}
Normalizing the warp factor to unity at the orbifold fixed point
$y = 0$, we get:
\be c_1 = {1 + \sqrt{1 -  {\omega}^2}}~. \label{c1}\ee
(The other solution $c_1=1-\sqrt{1-\omega^2}$, for which the RS
result is not recovered in the $\omega^2 \rightarrow 0$ limit, is
excluded from further discussions).

Next, to solve the hierarchy problem, we equate the warp factor at
$y=r\pi$ to the ratio of the Higgs to the Planck mass:
\be \label{ds} e^{-A} = \omega \cosh\left(\ln \frac {\omega}{c_1}
+ k r \pi \right) = 10^{-n}~.
 \ee
At this point, we keep $n$ arbitrary, although eventually we will
assume it to be $\simeq 16$. Defining $kr\pi\equiv x$, the above
equation simplifies to:
\be 10^{-n} = \frac{1}{2} \left[ c_1 e^{-x} + \frac{\omega^2}{c_1}
e^x \right] ~,\label{index11} \ee
from which one gets:
\be e^{-x} = \frac{10^{-n}}{2} \left[ 1 \pm \sqrt{1- \omega^2
10^{2n}} \right]~. \label{solnforx} \ee
Clearly, real solutions for $e^{-x}$ exists if and only if
$\omega^2 \leq 10^{-2n}$. In other words, solution of the
hierarchy problem requires the magnitude of the induced cosmological constant on
the brane to be extremely small! Thus from Eq.(\ref{c1}), $c_1
\simeq 2$, which we will assume from now on. Further setting the
brane cosmological constant $\omega^2 \equiv 10^{-N}$,
we get the upper bound on the cosmological constant:
\be N_{min} = 2n ~.\ee
Thus, for $n=16$, it follows that the brane cosmological constant
cannot exceed $10^{-32}$ (in Planck units).

Also Eqs.(\ref{index11}) and (\ref{solnforx}) simplify to:
\begin{eqnarray}
10^{-N} &=& 4 \left( 10^{-n} e^{-x} - e^{-2x} \right)~, \label{indexindex} \\
e^{-x} &=& \frac{ 10^{-n} }{2} \left[ 1 \pm \sqrt{1-10^{-(N-2n)} }
\right] ~. \label{2solns} \end{eqnarray}
From Eq.(\ref{2solns}) above, it can be seen that for $N
\rightarrow \infty$ ($\omega^2 \rightarrow 0$), the RS value of
$x=n\ln 10$ is recovered (the other solution in this limit,
$x=\infty$, is excluded). For $N=N_{min}$, we get a degenerate
solution $x=n\ln 10 + \ln 2$.
However, for $N<N_{min}$, there are {\it two} values of $x$ which
give rise to the required warping, instead of {\it one}, as was in
the case of RS \footnote{We thank A. Sen for pointing this out to
us.}. For $N-2n \gg 1$, these two solutions are:
\be x_1 \simeq n\ln 10 + \frac{1}{4}10^{-(N-2n)}~,~~ x_2 \simeq
(N-n) \ln 10 + \ln 4~. \ee
%

The first corresponds to the RS value plus a minute correction,
while the second, although of a similar order of magnitude (thus
ensuring that no new scale is introduced), is quite distinct. The
hierarchy problem is solved for two small and negative values of
the cosmological constant. Note that $x_2>x_1$.

In Figs.(\ref{ADSDS}) and (\ref{DSlarge}), we have plotted $N$
versus $x$, using Eq.(\ref{indexindex}), we have plotted $-N$
versus $x$ (for $n=16$). In Fig.(\ref{ADSDS}), point {\bf A}
corresponds to the RS values of $(x,N)=(n\ln 10,\infty)$. Point
{\bf B} corresponds to the maximum value of $\omega^2$, i.e.
$(x,N)=(n\ln 10 + \ln 2, 2n)$, beyond which $\omega^2$ starts to
decrease once again. Far from the maximum, $N$ is given
approximately by the linear relation:
\be N = \left( \ln 10 \right)^{-1} \left[ - x - n - \ln 4\right]~,
\ee
which we have plotted in Fig.(\ref{DSlarge}).

\begin{center} \begin{figure} \hfill
\begin{minipage}{.48\textwidth}
\includegraphics[width=3.330in,height=2.20in]{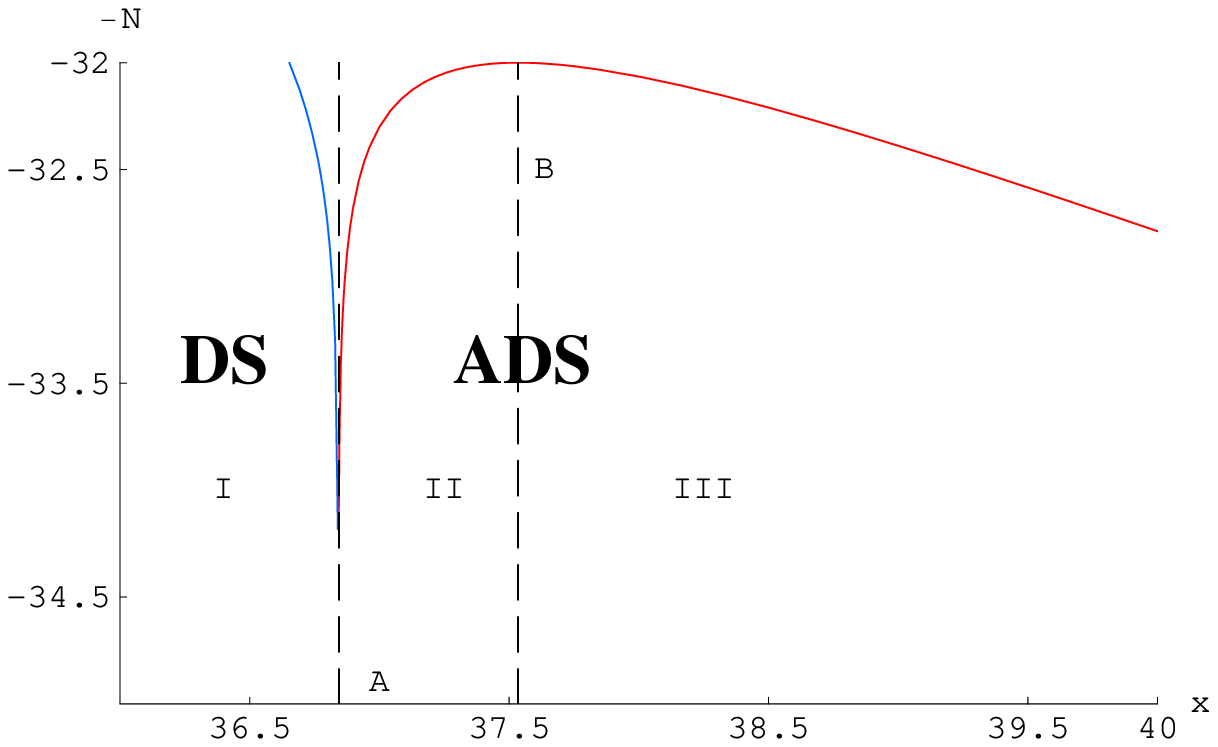}
\caption{Graph of $N$ versus $x=36-40$, for $n=16$ and for both
positive and negative brane cosmological constant. The curve in
region-I corresponds to positive cosmological constant on the brane,
whereas the curve in regions-II \& III represents negative
cosmological constant on the brane. } \label{ADSDS}
\end{minipage} \hfill \begin{minipage}{.48\textwidth}
\includegraphics[width=3.3030in,height=2.20in]{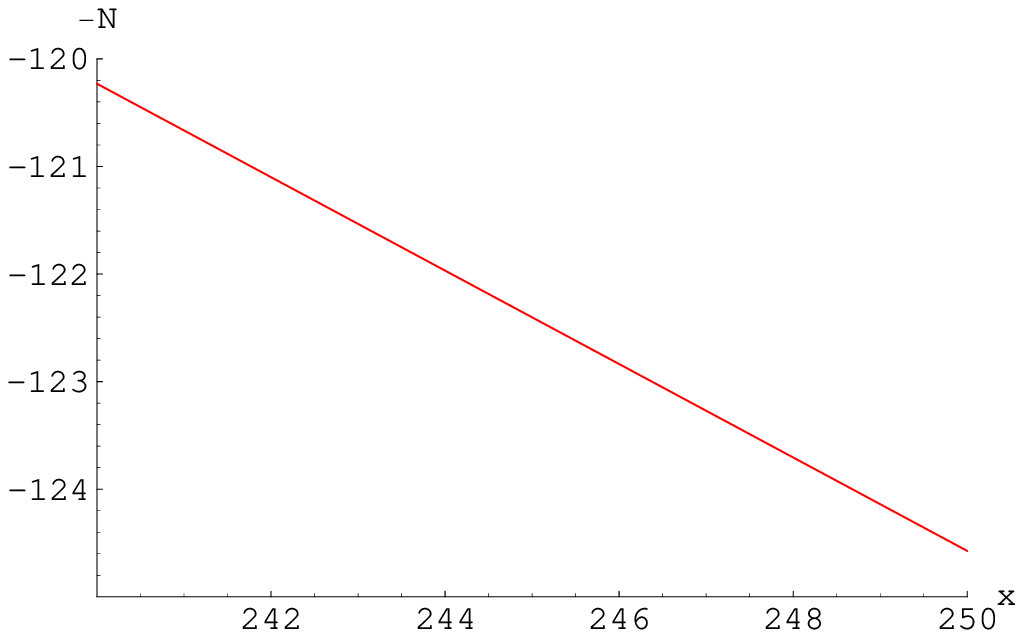}
\caption{(Continuation of of graph in Fig. 1) Graph of $N$ versus
$x=240-250$, for $n=16$ and negative brane cosmological constant}
\label{DSlarge} \end{minipage} \hfill \end{figure} \end{center}


From Eqs.(\ref{boundary1}) and (\ref{2solns}), we obtain the
tension on the visible brane for the two solutions as:
\be
{\cal V}_{vis} = (12 M^3 k)~
\frac{1 - 10^{N-2n} \left[ 1 \pm \sqrt{1- 10^{-(N-2n)}} \right] }
{ 1 + 10^{N-2n} \left[ 1 \pm \sqrt{1- 10^{-(N-2n)}} \right] }~.
\ee
Observe that ${\cal V}_{vis}=0$ when $N=N_{min}=2n$. Further, it
is easy to show that ${\cal V}_{vis} <0$ for $x=x_1$, while ${\cal
V}_{vis} > 0$ for $x=x_2$. Thus the second  solution for $x$ is
associated with a positive tension brane, which also produces the
desired large hierarchy. When $N-2n \gg 1$, the two tensions are
approximately given as:
\begin{eqnarray}
{\cal V}_{vis-1}
&\simeq& -(12M^3k) \\
{\cal V}_{vis-2} &\simeq& \frac{1}{3}(12M^3k)~. \label{tension4}
\end{eqnarray}
>From (\ref{tension4}), we see that a small negative cosmological constant
suffices to render the tension positive, provided the distance
between the branes is somewhat larger than the value predicted by
RS. From (\ref{boundary1}), the tension on the hidden brane on the
other hand, is given by:
\be {\cal V}_{hid} =  (12M^3k)~\frac{4-10^{-N}}{4 + 10^{-N}}~, \ee
which is always positive.


Next, for $\Omega > 0$, the warp factor which satisfies
Eqs.(\ref{eom3}-\ref{eom4}) is given by:
\be
 e^{-A} = \omega \sinh\left(\ln \frac {c_2}{\omega} - ky \right)~,
\ee
where now $\omega^2 \equiv \Omega/3k^2$, and as before,
normalization of the warp factor on the hidden brane gives $c_2 =
{1 + \sqrt{1 + {\omega}^2}}~.$

Equating the above to $m/m_0=10^{-n}$, we get:
\bea 10^{-n} = \frac{1}{2} \left[ c_2 e^{-x} -
\frac{\omega^2}{c_2} e^x \right] ~,\label{index1} \eea and the
counterpart of Eq.(\ref{solnforx}) is now,
\bea
%
%
e^{-x} &=& \frac{10^{-n}}{c_2} \left[ 1 + \sqrt{1+ \omega^2
10^{2n}} \right]~\label{solnforx2}.\eea
From Eq.(\ref{solnforx2}) one can see that in this case, there are
no bounds on $\omega^2$, and the (positive) cosmological constant
can be of arbitrary magnitude. Also, there is a single solution of
$x$, whose value depends on $\omega^2$ and $n$. This is described
in the region I in FIG.1, from where it can be seen that a small
and positive value of the cosmological constant, say the observed
value $\sim 10^{-124}$ (in Planckian units), corresponds to $x$
and hence $kr\pi$ very close to the RS value $36.84$ and the value of the cosmological constant rises
sharply with small departure from the RS value of $kr\pi$. This explains why the observed small value of cosmological constant
naturally leads to the tuning of the value of the modulus $r$ to be inverse of Planck length when the value of $k$ is
of the order of Planck mass. However in
this regime, the Tev brane tension continues to be negative as in
the RS case. This can be seen from the expressions for the brane
tensions which in this case are: \begin{equation} {\cal V}_{vis} =
12 M^3 k \left[\frac {\frac { \omega^2} {c_2^2} e^{ 2 k r \pi} +
1} { \frac { \omega^2} {c_2^2} e^{ 2 k r \pi} - 1}\right] ~ ;~
{\cal V}_{hid} =  12 M^3 k \left[\frac {1 + \frac { \omega^2}
{c_2^2}} { 1 - \frac { \omega^2} {c_2^2}}\right]
\label{boundary2}
\end{equation}
As $c_2 > \omega$, ${\cal V}_{hid}$ is always positive. On the
other hand from Eq.(30), the condition of positivity of the warp
factor $10^{-n}$ requires $\frac { \omega^2} {c_2^2} e^{ 2 k r
\pi} < 1$. This implies ${\cal V}_{vis}$
is negative for the entire range of positive values of $\Omega$.\\
In summary, we have derived the exact form for the warp factor in
a generalized RS braneworld scenario, which admits of both
positive or negative cosmological constant on the visible 3-brane.
We have shown that the induced negative cosmological constant on
the 3-brane is bounded from below by $\sim -10^{-32}$ (in Planck
units). Furthermore for a tiny value of cosmological constant, the
hierarchy problem can be resolved for two different values of the
modulus, one of which corresponds to a positive tension Tev brane
along with the positive tension Planck brane. It would be
interesting to study implications of these results. In the other
regime namely $\Omega > 0$ the Tev brane tension turns out
necessarily to be negative. The value of the modulus corresponding
to the observed value of the cosmological constant lies very close
to the RS value and rises very rapidly as we depart from the value of $kr$ predicted in the original RS model.
Tuning of the small cosmological constant thus implies the  tuning of the value of the modulus $r$ at the inverse of
Planck length. It may be noted that a modulus value much away from Planck length ( corresponding to a large
cosmological constant ) will give rise to
a new hierarchy of scales leading to a possible large radiative correction to the modulus and in turn
bringing back the fine tuning problem again. 
Our results thus indicate that if one wants to
resolve the fine tuning problem in connection with the Higgs mass,
without bringing in any hierarchy through the size of the modulus,
the value of the cosmological constant $\Omega$ (whether positive
or negative) on the Tev brane must be very small! In other words
the resolution of the gauge hierarchy problem, and the cosmological fine
tuning problem are related and one implies the other if the modulus value is kept close to Planck length to avoid any further scale
hierarchy. 
It will now be interesting to study whether for this generalised RS model 
the modulus can be stabilized to a value close to Planck length  following the mechanism proposed by Goldberger and Wise \cite{goldwise} and
what are the other phenomenological/cosmological implications of such a generalised warp factor and brane tensions. 
We hope to report these in  future works.\\
We thank A. Dasgupta, A. Sen and S. Sur for useful discussions. SD
thanks the Department of Theoretical Physics, Indian Association
for the Cultivation of Science, Kolkata, for hospitality, where
this work was done. This work is supported by the Natural Sciences
and Engineering Research Council of Canada.


\end{document}